\RequirePackage{fix-cm}

\documentclass[twocolumn]{svjour3}          

\smartqed  
\usepackage{graphicx}
\usepackage{setspace}
\usepackage[intlimits]{amsmath} 
\usepackage{bm}
\usepackage{txfonts}

\begin{document}

\title{Generalized proof of uncertainty relations in terms of commutation relation and interpretation based on action function}

\author{Chol Jong$^{1}$\and Shin-Hyok Jon$^1$\and Son-Il Jo$^1$\and Nam-Chol Choe$^1$}

\institute{Chol Jong\\\email{jch59611@star-co.net.kp} \\ \\
Nam-Chol Choe\\\email{cnc81103@star-co.net.kp} \\ \\
$^1$ Faculty of Physics, Kim Chaek University of Technology, Yonggwang Street, Pyongyang, Democratic People's Republic of Korea\\
}

\date{Received: date / Accepted: date}

\maketitle

\begin{abstract}

The uncertainty principle is the most important for the foundations of quantum mechanics but it still remains failed to reach a consensus of its interpretation, which gives rise to debates upon its physical nature.
In this work, we address the problem of its foundation from a different aspect to present an alternative formulation for proving the uncertainty relations in a general way in terms of commutation relations and action function.
The relationship between the de Broglie relation and the uncertainty principle is studied from a new angle.
As a result, it is demonstrated that the de Broglie relation is the foundation of the uncertainty principle.
Starting with the origin of the problem, we show that the de Broglie relation provides the form of the wave function and the determined form of the wave function in turn leads to the conception of operators for quantum mechanics, and thus it is possible to provide with the help of the operators and wave function a generalized proof of the uncertainty principle as the law governing ensemble of states.
As a decisive solution to the problem, the interpretation of the uncertainty principle in terms of the action function is offered that gives one self-consistent explanation in agreement with the known physical phenomena.
Eventually, we show the necessity and possibility of reassessing and improving the foundation and interpretation of the uncertainty principle as a leading principle of quantum mechanics.  
\end{abstract}


\section{Introduction}\label{intro}

The uncertainty principle is of an important significance in the foundations and interpretation of quantum mechanics.
There is no exaggeration in saying that the formulations and foundations of quantum mechanics are to be fundamentally distinguished by the interpretation of this principle. 
The uncertainty principle involves the laws related to simultaneous measurements of canonically conjugate dynamical quantities, i.e., several uncertainty relations.
For each pair of canonically conjugate variables, the product of the uncertainties given by the measurement of both quantities cannot be less than a fixed value, i.e., $\hbar /2$. 
It is known that the uncertainty relation is established for position and momentum, time and energy, and angle and angular momentum.
The uncertainty relation for energy and time can be written in a form identical to that for position and momentum, but
there is the essential difference between the two uncertainty relations.
That is due to the fact that the time and energy are not operators  \cite{Auletta}.  
The uncertainty relation for the angular momentum and angle is similar to that for the energy and time because the angular momentum operator is not known so far.
The analysis of the uncertainty relation for the angular momentum and angle was performed by Judge \cite{Judge}.
Another uncertainty relation which is of particular importance in quantum optics exists as the relation between the number of photons expressed by the number operator and the phase \cite{Louisell}. 
In addition, there is the uncertainty relation for the spin and spin phase which makes use of the positive operator valued measure \cite{Busch}.

Although the uncertainty principle seems to be justifiable, the long history of its interpretation has spawned a multitude of different opinions \cite{Auletta,Heisenberg1,Heisenberg2,Heisenberg3,Bohr,Einstein,Schrodinger,Landau,Fock1,Fock2,Mandelstam,Hilgevoord1}.
Heisenberg considered the uncertainty relations to be experimental consequences of the measuring processes \cite{Heisenberg1,Jensen,Jordan}.
His argument affirms the maximum attainable certainty product, i.e., the minimum uncertainty product allowed by the uncertainty principle. 
Heisenberg identified the uncertainty with the standard deviation of measured observables. 
In his view, the uncertainty principle expresses a type of disturbance originated from the interaction between apparatus and quantum system.  
Heisenberg's analysis itself was rather qualitative in character. 
The first form of explicit mathematical derivation of the analysis had been given by Kennard \cite{Kennard}. 

Robertson, Schr\"{o}dinger and Beltrametti proposed the statistical formulations of the uncertainty principle \cite{Auletta,Schrodinger,Robertson,Beltrametti}.
They proved the uncertainty relation for arbitrary conjugate observables by determining the product of the square roots of the statistical deviation from mean value of the observables in terms  of operators and wave function. 
Although Robertson's formulation is commonly used, it is somehow unsatisfactory \cite{Deutsch}.
Bohr interpreted the uncertainty principle as the complementary principle based on particle-wave dualism which is distinguished from that of Heisenberg \cite{Bohr,Rosenfeld}.
What is very interesting in all Bohr's thought experiment is that there is absolutely no mention of the standard deviation or of the variance \cite{Hilgevoord1}. 
The recent researches into the uncertainty principle tend to intensify\cite{Maassen,Kraus,Schurmann,Sen,Fedullo,Rastegin}.
Entropic uncertainty relations were proposed as a useful alternative approach \cite{Bialynicki1,Portesi,Wehner,Bialynicki2}. 

For all efforts, the uncertainty principle  still remains unsolved in many respects.
Above all, it is not clear what the uncertainty is.
Despite very wide prevalence of the uncertainty principle, there is no general consensus over its scope and validity \cite{Hall,Busch2}. 
Whether the uncertainty is due to the standard deviation or the dualism of material or the ensemble of states still remains an open question.

In the next place, we so far cannot prove several uncertainty relations in a unique way, for example, in terms of the commutation relations of canonically conjugate observables.
There are observables in quantum mechanics which are not associated with operators: energy and time, and 
angle and angular momentum \cite{Auletta}.
In relativistic quantum mechanics, there does not exist a position operator for the photon.
Therefore, we cannot equally apply the commutation relation between canonically conjugate observables to prove the uncertainty relations.  
On the other hand, it should be noted that complementary observables are not always a pair. 
Actually, for every observable, it is possible to find many operators which do not commute with it \cite{Neumann}.
In this connection, it is natural to raise the question of how to interpret the uncertainty relation between them.
Meanwhile, for a certain observable constituted by  two observables that are canonically conjugate we cannot justify the uncertainty principle because the measurement of such a observable itself is not allowed due to the impossibility of simultaneous measurement.
As a result, the uncertainty principle extends the impossibility of measurement even to an observable beyond a pair of canonically conjugate observables.
As an example, we can take the angular momentum.

This fact shows that the foundation of the uncertainty principle is unclear.
Evidently, it still remains an interesting question to construct the airtight and general formulation of the uncertainty principle.

In this work, we aim to prove the uncertainty principle in a general way in terms of the commutation relations and to reveal the foundation of the uncertainty principle. 

\section{Possibility of simultaneous determination of position and momentum}\label{Possibility}

According to the standard theory of quantum mechanics,  choosing a position and the corresponding canonically conjugate momentum variable simultaneously is regarded as violating the uncertainty principle.
On the contrary, our view is that taking both position and momentum variable as basic variables of a quantum-mechanical state does not invoke any inconsistency.
In quantum mechanics, positions of particles are variables for indicating the probability of  finding particles in a volume element in the vicinity of a given position of configuration space. 
From the point of view of statistical interpretation, it is obvious that the Schr\"{o}dinger equation deals with ensembles in configuration space for positions of particles. 

The standard theory of quantum mechanics excludes the possibility of the simultaneous determination of position and momentum according to the uncertainty principle.
However, it  is necessary to deliberate on the fact that the application of the momentum operator to a wave function definitely determines momenta of particles.
In fact, a momentum operator makes a position uniquely correspond  to a definite momentum via a wave function in configuration space.
The correspondence refers to $ \mathbf{q}~\rightarrow~ \psi(\mathbf{q})~\rightarrow\hat{\mathbf{p}}\psi(\mathbf{q})~\rightarrow\mathbf{p}$ attained with the help of the momentum operator.
Therefore, it is obvious that for configuration space the introduction of the momentum operator means the one-to-one correspondence between position and momentum.
However, it should be considered that in phase space there is not the one-to-one correspondence but multi-correspondence relation between positions and momenta. 
Nevertheless, the simultaneous determination of position and momentum is not problematic, since it is ensured by quantum operators and wave function.
In fact, the momentum operator uniquely determines momenta via the wave function.
Of course, the consideration in such a direction certainly is inconsistent with the uncertainty principle. 

Also, there is other logic to adduce reason in support of the possibility of the simultaneous determination of position and momentum. 
It is necessary to give careful thought to the fact that the velocity of a particle is calculated by the time series of positions, and the momentum is determined as the product of mass and velocity.
Meanwhile, according to the standard quantum theory, it is possible to exactly measure without limitation both time and position simultaneously, since they are not canonically conjugate.
This possibility therefore makes us circumvent the uncertainty principle  to determine the position and momentum simultaneously as exactly as we want.

In principle, there is another way which makes us avoid the disturbance which the simultaneous measurement of position and momentum causes. 
If we make not simultaneous but alternate measurement of position and momentum, according to the uncertainty principle the results of measurement prove to be free of disturbance.
Then the data on position and momentum obtained by the alternate measurement become non-perturbed information reflecting the reality of a microscopic particle.
In this case, it is obvious that the position and the momentum function interpolated on the basis of the time series of position and momentum observables should be identified with each other by derivation or integration.
It shows that position and momentum have correspondence relation and the alternate and simultaneous measurements are equivalent.

Heisenberg's interpretation of the uncertainty relation is based on the disturbance that the joint measurement of canonically conjugate observables brings about for each other.
If the uncertainty of momentum is ascribed to the disturbance due to the measurement of position, then we should regard  the sequential measurement of position itself as being disturbed by previous measurements of position.
If so, it is impossible to study quantum phenomena with exactly measured values of observables.
The wave function and operators as the starting point for investigating the uncertainty relation do not embody the measurement by observer at all and purely represent the probabilistic possibilities inherent in quantum system. 
Therefore, the uncertainty principle has nothing to do with measurement.  

This logic of the simultaneous determination of position and momentum can be represented by the following diagram.

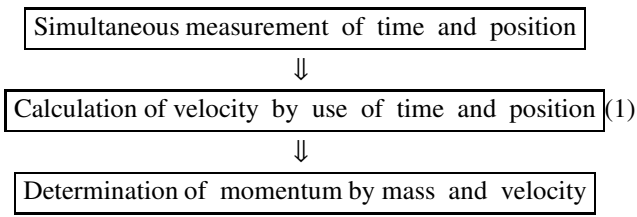
\begin{eqnarray}\label{eq:1}
&\framebox{Simultaneous~measurement~ of~ time~ and~ position} \nonumber\\
&\Downarrow\nonumber\\
&\framebox{Calculation~of~velocity~ by~ use~ of~ time~ and~ position} \\ 
&\Downarrow\nonumber\\
&\framebox{Determination~of~ momentum~by~mass~ and~ velocity}\nonumber
\end{eqnarray}

It is worth noting that until the sixties it was 
a generally believed that it is impossible to measure two conjugate observables together, but since then the matter has not been the case; it is currently accepted that it is possible to measure two conjugate observables together \cite{Auletta}.
The emergence of the theory of the positive-operator valued measurement (POVM) is a significant example that shows the renovation in the understanding of the uncertainty principle  \cite{Busch,Muynck}. 

Judging from this ground, we are able to imagine a definite set of positions and the corresponding momenta.  
In this context, an ensemble in position space definitely corresponds to that in momentum space.

The main problem is to explain how these two distributions are related.
Of course, this relation can be rigorously proved rather than assumed, thereby showing that the uncertainty relation is derived by treating the statistical ensemble. 
In this work, we aim to prove the reciprocal relation between the distribution of position and that of momentum whose product should have a fixed minimal value rather than zero.

Quantum states can be described by ensembles which obey the wavelike rule.
The quantum ensemble displays probabilistic property which characterizes the wave field expressed as the correlation between phase trajectories, i.e., self-interference. 
The correlation is governed by the de Broglie relation, i.e., the wavelike property. 
The quantum ensemble therefore should assume both the dynamical laws for particles and the wavelike property of the phase-space ensemble characterizing the wave field. 
This ensemble determines the distribution in phase space, while the distribution is supposed to bear the reciprocal relation between canonically conjugate quantities  referred to as the uncertainty relation. 
We shall prove this reciprocal relation based on the commutation relations between operators and alternatively based on the action function.

The unavoidable one-to-one correspondence between position and momentum demonstrated purely by the standard theory of quantum mechanics shows that it is necessary to renew our understanding of the uncertainty relation. 
For this reason, we cannot but adopt definitely phase space even for microscopic world.

\section{Action function as foundation of operators}\label{foundation}

It is important to demonstrate that the de Broglie relation suggests the form of the wave function.

The explanation runs as follows.
It is obvious that the de Broglie's relation defines the frequency and the wave vector of the de Broglie wave.
Using this relation, we can determine the phase of the wave without loss of generality as the function with variable upper limits, $\mathbf{q}, t$
\begin{equation}\label{eq:2}
\Phi=\int_0^{\mathbf{q}} \mathbf{k}(\mathbf{q}')~ \mathrm{d}\mathbf{q}' -\int_0^t  \omega(t') \mathrm{d}t'.
\end{equation}
Of course, the  phase of a free particle is represented as
\begin{equation*}
\Phi= \mathbf{k}~\mathbf{q} -\omega t.
\end{equation*}
For a period, the phase relation should satisfy
\begin{equation}\label{eq:3}
\int_0^{\mathbf{q}} \mathbf{k}(\mathbf{q}')~ \mathrm{d}\mathbf{q}' -\int_0^t  \omega(t') \mathrm{d}t'=2\pi.
\end{equation}
The above relation indicates the minimal condition necessary for a wave.
In order for a wave to exist, it should satisfy at least this physical condition.
According to the de Broglie relation: 
\begin{equation*}
\mathbf{p}=\hbar \mathbf{k},~~~ E=\hbar\omega, 
\end{equation*}
we get from Eq. (\ref{eq:3})
\begin{equation}\label{eq:4}
\int_0^{\mathbf{q}} \mathbf{p}(\mathbf{q}')~ \mathrm{d}\mathbf{q}' -\int_0^t  E(t') \mathrm{d}t'=2\pi \hbar=h.
\end{equation}
Here, we supposed that even though the momentum of a particle changes, the de Broglie relation is applicable as in the case of a free particle.
Hence, we can adopt the general condition of periodicity as
\begin{equation*}
\int_0^{\mathbf{q}}  \mathbf{p}(\mathbf{q}')~ \mathrm{d}\mathbf{q}' -\int_0^t  E(t') \mathrm{d}t'=nh,
\end{equation*}
where $n$ is an integer.
The condition of periodicity of a free particle is represented as
\begin{equation*}
\mathbf{p}\mathbf{q} -Et=nh.
\end{equation*}
This is nothing but the Bohr-Sommerfeld quantization condition.
Consequently, we in general can write the phase of the de Broglie wave as
\begin{equation}\label{eq:5}
\Phi=2\pi\frac{\int_0^{\mathbf{q}}  \mathbf{p}(\mathbf{q}')~ \mathrm{d}\mathbf{q}' -\int_0^t  E(t') \mathrm{d}t'}{h}.
\end{equation}
Here, $\int_0^{\mathbf{q}} \mathbf{p}(\mathbf{q}')~ \mathrm{d}\mathbf{q}' -\int_0^t  E(t') \mathrm{d}t'$ obviously is the action function with variable upper limit, which is represented in terms of the Hamilton function as
\begin{equation}\label{eq:6}
S\left(\mathbf{q}, \mathbf{p}, t\right)=\int_0^{\mathbf{q}} \mathbf{p}\left(\mathbf{q}'\right)~ \mathrm{d}\mathbf{q}' -\int_0^t H(t') \mathrm{d}t'.
\end{equation}
That momentum is qualified as an independent variable is due to the introduction of ensemble of paths.
A variety of virtual paths that represents the probabilistic behaviour of microscopic particles produces a spectrum of momenta.
Therefore, the action function must be defined in phase space.

From Eq. (\ref{eq:6}), the action function of a free particle is 
\begin{equation*}
S\left(\mathbf{q}, \mathbf{p}, t\right)=\mathbf{p}~\mathbf{q} -H t.
\end{equation*}

Consequently, the phase of the probability wave is written in terms of the action function as
\begin{equation}\label{eq:7}
\Phi=\frac{S\left(\mathbf{q}, \mathbf{p}, t\right)}{\hbar}.
\end{equation}
In the end, the de Broglie wave can be represented as 
\begin{equation}\label{eq:8}
\psi\left(\mathbf{q},\mathbf{p},t\right)=\varphi(\mathbf{q},\mathbf{p},t) \exp \left(\mathrm{i}\frac{S\left(\mathbf{q},\mathbf{p},t\right)}{\hbar}\right),
\end{equation}
where $\varphi\left(\mathbf{q},\mathbf{p},t\right)$ as the probability amplitude is a real-valued function.

We therefore can conclude that the de Broglie relation enables us to determine the form of the wave function.
Meanwhile, Eq. (\ref{eq:4}) shows essential contents of the uncertainty relation reflecting ensemble in phase space and the broad context of classical mechanics and quantum mechanics.
Expression (\ref{eq:4}), (\ref{eq:5}) tell us that the quantum of the action is $h$.
At the same time, it shows the necessity and validity of the simultaneous determination of position and momentum, and of time and energy.
This is because if it were not to be possible, we could not imagine the phase of a wave.

Of course, this assumption cannot be regarded to be new, since such a form of wave function has already been used in the preceding formulations.
It is necessary to recall the fact that the Schr\"{o}dinger equation was obtained,  implicitly employing this assumption.
In fact, for the  Schr\"{o}dinger equation the phase part of the wave function assumed for a free particle is in accordance with this assumption. 
Such an understanding of the wave field serves as the basis for establishing  the present formalism of quantum mechanics in phase space.

We can suppose from Eq. (\ref{eq:4})
that the quantum in phase space should be represented as
\begin{equation}\label{eq:9}
h=\int_0^{\mathbf{q}} \mathbf{p}(\mathbf{q}')~ \mathrm{d}\mathbf{q}',
\end{equation}
while the quantum in energy-time space should be represented as
\begin{equation}\label{eq:10}
h=\int_0^t H(t') \mathrm{d}t'.
\end{equation}
These relations really shed light on the nature of space-time quantization.
It is natural to interpret these relations as characterizing an ensemble that consists of pairs of position and momentum, and of time and energy.
Relations (\ref{eq:9}), (\ref{eq:10}) show that the greater the momentum is, the more the space is localized, while the greater the energy is, the shorter the time of quantum process.

The statistical formalism of quantum mechanics enables us to conceive operators as the tool for deriving dynamical quantities from the wave function. 
The application of a differential operator to wave function should yield the corresponding dynamical quantity.
In the end, we come to find out the operator relations equal or analogous to ones in the Schr\"{o}dinger equation.

What should be stressed is that the obtained results are thanks to the form of wave function, Eq. (\ref{eq:8}).
Of course, from the point of view of dynamics the assumption about the form of the wave function is explained in some degree by Ref. \cite{Jongchol}.
To begin with, we calculate the derivatives of the action function in the extended phase space with respect to $q, p, t$.

To advance further, we take the essential step as follows.
\begin{align}\label{eq:11}
&\frac{\partial S\left(\mathbf{q}, \mathbf{p}, t\right)}{\partial q_i}=\displaystyle\frac{\partial}{\partial q_i}\left( \int_0^{\mathbf{q}} \mathbf{p}(\mathbf{q'})\mathrm{d}{\mathbf{q'}} -\int_0^t H(t')\mathrm{d}t'\right)\nonumber\\
&=\displaystyle\frac{\partial}{\partial q_i}\left( \int_0^{\mathbf{q}} \mathbf{p}(\mathbf{q'})\mathrm{d}{\mathbf{q'}} \right)
=\frac{\partial}{\partial q_i}\left(\mathbf{p}(\mathbf{q'})\mathbf{q'}|^\mathbf{q}_0 -\int_0^{\mathbf{q}} \mathbf{q'}\frac{\partial\mathbf{p}(\mathbf{q'})}{\partial\mathbf{q'}}\mathrm{d}{\mathbf{\mathbf{q'}}} \right)\nonumber\\
&=\displaystyle\frac{\partial}{\partial q_i}\left(\mathbf{p}(\mathbf{q'})\mathbf{q'}|^\mathbf{q}_0 \right)=\frac{\partial}{\partial q_i}\left(\mathbf{p}\mathbf{q} \right)=p_i,
\end{align}

\begin{align}\label{eq:12}
&\frac{\partial S\left(\mathbf{q}, \mathbf{p}, t\right)}{\partial p_i}=\frac{\partial}{\partial p_i}\left( \int_0^{\mathbf{q}} \mathbf{p}(\mathbf{q'})\mathrm{d}{\mathbf{q'}} -\int_0^t H(t')\mathrm{d}t'\right)\nonumber\\
&=\displaystyle\frac{\partial}{\partial p_i}\left( \int_0^{\mathbf{q}} \mathbf{p}(\mathbf{q'})\mathrm{d}{\mathbf{q'}} \right)
=\frac{\partial}{\partial p_i}\left(\mathbf{p}(\mathbf{q'})\mathbf{q'}|^\mathbf{q}_0 -\int_0^{\mathbf{q}} \mathbf{q'}\frac{\partial\mathbf{p}(\mathbf{q'})}{\partial\mathbf{q'}}\mathrm{d}{\mathbf{\mathbf{q'}}} \right)\nonumber\\
&=\displaystyle\frac{\partial}{\partial p_i}\left(\mathbf{p}(\mathbf{q'})\mathbf{q'}|^\mathbf{q}_0 \right)=\frac{\partial}{\partial p_i}\left(\mathbf{p}\mathbf{q} \right)=q_i,
\end{align}
\begin{equation}\label{eq:13}
\frac{\partial S\left(\mathbf{q}, \mathbf{p}, t\right)}{\partial t} =\frac{\partial}{\partial t}\left(-\int_0^tH(t')dt'\right)=-H.
\end{equation}

Next, let us review $\mathrm{\mathrm{i}}\hbar \displaystyle\frac{\partial \psi}{\partial t}$.
The application of $\mathrm{i}\hbar \displaystyle\frac{\partial}{\partial t}$ to the wave function yields
\begin{eqnarray}\label{eq:14}
\mathrm{i}\hbar \frac{\partial \psi}{\partial t}=\mathrm{i}\hbar \frac{\mathrm{i}}{\hbar}\frac{\partial S}{\partial t}\:\psi+\mathrm{i}\hbar \frac{1}{\varphi}\frac{\partial \varphi}{\partial t}\:\psi
=H \psi+\mathrm{i}\hbar \frac{1}{\varphi} \frac{\partial \varphi}{\partial t}\:\psi.
\end{eqnarray}
From the above expression, we can interpret $\mathrm{i}\hbar \displaystyle\frac{\partial}{\partial t}$ as the operator relative to total energy, since it makes $H$ be derived from the wave function.

Furthermore, let us examine $-\mathrm{i}\hbar \displaystyle\frac{\partial \psi}{\partial p_i}$.
The application of this operator to the wave function produces
\begin{eqnarray}\label{eq:15}
-\mathrm{i}\hbar \frac{\partial \psi}{\partial p_i}=-\mathrm{i}\hbar \frac{\mathrm{i}}{\hbar}\frac{\partial S}{\partial p_i}\:\psi-\mathrm{i}\hbar \frac{1}{\varphi}\frac{\partial \varphi}{\partial p_i}\:\psi
=q_i \psi-\mathrm{i}\hbar \frac{1}{\varphi} \frac{\partial \varphi}{\partial p_i}\:\psi.
\end{eqnarray}
Since this operation gives $q_i$, we can regard $-\mathrm{i}\hbar \displaystyle\frac{\partial}{\partial p_i}$ as the position operator.

Similarly, we have
\begin{eqnarray}\label{eq:16}
-\mathrm{i}\hbar \frac{\partial \psi}{\partial q_i}=-\mathrm{i}\hbar \frac{\mathrm{i}}{\hbar}\frac{\partial S}{\partial q_i}\:\psi-\mathrm{i}\hbar \frac{1}{\varphi}\frac{\partial \varphi}{\partial q_i}\:\psi
=p_i \psi-\mathrm{i}\hbar \frac{1}{\varphi} \frac{\partial \varphi}{\partial q_i}\:\psi.
\end{eqnarray}
As a consequence, $-\mathrm{i}\hbar \displaystyle\frac{\partial}{\partial q_i}$ is adopted as the momentum operator.
The obtained results naturally bring the idea of operator to us. 
From Eqs. (\ref{eq:14}), (\ref{eq:15}), (\ref{eq:16}), we can interpret the meaning of the relation between an observable, $L$ and the corresponding operator, $\hat{L}$ as 
\begin{equation}\label{eq:17}
L=\mathrm{Re} \left( \frac{  \hat{L}\psi }{\psi}\right).
\end{equation}
In fact, this relation naturally comes from the definition of mean value.
By definition, the mean value with respect to $\hat{L}$ is written as
\begin{equation}\label{eq:18}
\bar{L}=\int \psi^\ast \hat{L} \psi \mathrm{d}\tau=\int \psi^\ast \psi \frac{\hat{L} \psi}{\psi} \mathrm{d}\tau=\int \left|\psi\right|^2 \frac{\hat{L} \psi}{\psi} \mathrm{d}\tau.
\end{equation}
Accordingly, $\mathrm{Re} \left( \displaystyle\frac{\hat{L} \psi }{\psi}\right)$ should be regarded as the observable with respect to operator $\hat{L}$.

By inference, we find the operators corresponding to kinetic energy and potential. 
Altogether, the operators corresponding to fundamental observables are represented as

\begin{align}
&\hat{E}=\mathrm{i}\hbar \frac{\partial}{\partial t},\label{eq:19}\\
&\hat{p_i}=-\mathrm{i}\hbar \frac{\partial}{\partial q_i},\label{eq:20}\\
&\hat{q}_i=-\mathrm{i}\hbar \frac{\partial}{\partial p_i},\label{eq:21}\\
&\hat{U}=-\frac{\mathrm{i}\hbar}{2} \sum_{i=1}^{f} -\frac{\partial H}{\partial q_i}\:\frac{\partial}{\partial p_i}= \frac{1}{2}\sum_{i=1}^{f} \left(\dot{p}_{group}\right)_i~\hat{q}_i,\label{eq:22}\\
&\hat{T}=-\frac{\mathrm{i}\hbar}{2}\sum_{i=1}^{f} \frac{\partial H}{\partial p_i} \frac{\partial }{\partial q_i}=\frac{1}{2} \sum_{i=1}^{f}\left(\dot{q}_{group}\right)_i\:\hat{p}_i\label{eq:23},
\end{align}
where $\left(\dot{q}_{group}\right)_i$ and $\left(\dot{p}_{group}\right)_i$ denote the $i$th components of the group velocity with respect to position and momentum, respectively \cite{Jongchol}. 
The successive three operators correspond to energy, momentum and position, respectively, which become basic dynamical quantities. 
The fourth operator should be considered to be the potential energy operator since it corresponds to a potential energy function. 
This operator suggests nothing but the virial theorem of statistical mechanics.
Thus, we can arrive at the important conclusion that in quantum mechanics the potential energy should be represented as the virial of the system under consideration.
Meanwhile, the fifth operator should be considered the kinetic energy operator, since it corresponds to kinetic energy. 

Since the action function, Eq. (\ref{eq:6})  is represented by generalized coordinates and generalized momenta, naturally the angular momentum operator should have the following form.
\begin{equation}\label{eq:24}
\hat{L}_{\varphi}=-\mathrm{i}\hbar \frac{\partial}{\partial \varphi}.
\end{equation}

The difference of the operators from ones in the Schr\"{o}dinger equation consists in the fact that the wave functions applied by them are defined in phase space. 
For the Schr\"{o}dinger equation, the wave function is the state function defined in configuration space, whereas for the fundamental equation of quantum mechanics in phase space the wave function is the state function defined in phase space.
 
It is emphasized that the dynamical quantities obtained with the help of operators and wave function are not the same as classical ones, and get quantal.

Such an interpretation of quantum observables naturally leads to adopting time as an ordinary quantum observable.
Extending the phase space furthermore, we can take the action function as
\begin{equation}\label{eq:25}
S\left(\mathbf{q}, \mathbf{p}, H, t\right)=\int_0^{\mathbf{q}} \mathbf{p}\left(\mathbf{q}'\right)~ \mathrm{d}\mathbf{q}' -\int_0^t H\left(t'\right) \mathrm{d}t',
\end{equation}
then we get
\begin{align}\label{eq:26}
&\frac{\partial S\left(\mathbf{q}, \mathbf{p}, H, t\right)}{\partial H}=\frac{\partial}{\partial H}\left( \int_0^{\mathbf{q}} \mathbf{p}\left(\mathbf{q}'\right)~ \mathrm{d}\mathbf{q}'-\int_0^t H(t')\mathrm{d}t'\right)\nonumber\\
&=-\frac{\partial}{\partial H}\left( \int_0^t H(t')\mathrm{d}{t'} \right)
=-\frac{\partial}{\partial H}\left(H(t')t'|^{t}_0 -\int_0^t t'\frac{\partial H(t')}{\partial t'}\mathrm{d}{t'} \right)\nonumber\\
&=-\frac{\partial}{\partial H}\left(H(t')t'|^t_0 \right)=-\frac{\partial}{\partial H}\left(H t \right)=-t.
\end{align}
The time operator therefore becomes 
\begin{equation}\label{eq:27}
\hat{t}=\mathrm{i}\hbar \frac{\partial}{\partial H}~.
\end{equation}
The time operator should be considered to apply to the wave function in phase space as follows.
\begin{equation}\label{eq:28}
\hat{t}=\mathrm{i}\hbar \frac{\partial}{\partial H}=\mathrm{i}\hbar \sum_i\left[\left(\frac{\partial H}{\partial q_i}\right)^{-1}\frac{\partial}{\partial q_i}+\left(\frac{\partial H}{\partial p_i}\right)^{-1}\frac{\partial}{\partial p_i}\right]+\left(\frac{\partial H}{\partial t}\right)^{-1}\frac{\partial}{\partial t}.
\end{equation}
Hereafter, we represent the time operator as
\begin{equation}\label{eq:29}
\hat{t}=\mathrm{i}\hbar \frac{\partial}{\partial E}.
\end{equation}
Ultimately, the special status of time as an exceptional observable without the corresponding operator comes to be lost and the system of basic operators of quantum mechanics becomes complete.

The introduction of these operators helps to clarify the relations of this formalism with the others.
Using the above operators, we can represent the fundamental equation of quantum mechanics in phase space \cite{Jongchol} as 
\begin{equation}\label{eq:30}
\hat{E}\psi=\frac{1}{2} \left\{\sum_{i=1}^{f} \left( \frac{\partial H}{\partial p_i}\hat{p}_i- \frac{\partial H}{\partial q_i} \hat{q}_i\right)\right\}\psi
\end{equation}
or
\begin{equation}\label{eq:31}
\hat{E}\psi=\frac{1}{2} \left\{\sum_{i=1}^{f} \left[ \left(\dot{q}_{group}\right)_i\hat{p}_i+\left(\dot{p}_{group}\right)_i\hat{q}_i\right]\right\}\psi.
\end{equation}
In more compact form, we write
\begin{equation*}
\hat{E} \psi= \hat{H} \psi~,
\end{equation*}
where in view of Eq. (\ref{eq:22}) and (\ref{eq:23}) the Hamiltonian operator takes the following form: 
\begin{equation}\label{eq:32}
\hat{H}= \hat{T}+\hat{U}.
\end{equation}
The fundamental equation of this formalism is distinguished from the Schr\"{o}dinger equation because the wave function is defined not in configuration space but in phase space. 
This formalism is expected to be useful to elucidate the essence of the uncertainty relations and to prove them generally. 

\section{Generalized proof of uncertainty relations in terms of commutation relations}

\subsection{Proof of uncertainty relation for position and momentum in terms of commutation relation}
\
\indent The statistical formalism of quantum mechanics enables us to prove the uncertainty relations in a general way in the sense of the standard deviation by means of the commutation relation.
The momentum operator assumed in this formalism, which applies to the wave function in phase space, is in exact accord with that in the standard theory of quantum mechanics. 
Therefore, it is concluded that for the two cases the commutation relation between position and momentum operator is identical. 
In fact, the uncertainty relation can be similarly proved for the two case. 

Now, we shall prove the uncertainty relation with regard to the wave function in phase space. 
Let the domain of variability of $p$ and $x$ be $p \in[p_1, p_2]$ and $x \in [x_1, x_2]$. 
As usual, we may take the variability as $p_1=-\infty, p_2=\infty, x_1=-\infty, x_2=\infty$. 
At the boundaries, the values of the wave function vanish.

With the help of the commutation relation between position and momentum operator
\begin{equation*}
\left(x \hat{p}-\hat{p}x \right) =-\mathrm{i} \hbar \left(x \frac{\partial }{\partial x}- \frac{\partial}{\partial x} x \right) =\mathrm{i} \hbar,
\end{equation*}
we calculate 
\begin{align}\label{eq:33}
&\displaystyle\int_{p_1}^{p_2} \mathrm{d}p \int_{x_1}^{x_2} \mathrm{d} x \left| \alpha x \psi + \frac{\partial \psi}{\partial x} \right|^2  \nonumber \\
&=\int_{p_1}^{p_2} \mathrm{d}p \int_{x_1}^{x_2} \mathrm{d} x \left( \alpha x \psi^\ast + \frac{\partial \psi^\ast}{\partial x}\right) \left( \alpha x \psi + \frac{\partial \psi}{\partial x} \right) \nonumber \\
&= A \alpha^2 + B \alpha + C \geq 0.
\end{align}
Hence, we find
\begin{equation*}
B^2-4AC \leq 0 \rightarrow \frac{B^2}{4} \leq AC.
\end{equation*}
Then the calculated $A$, $B$ and $C$ are as follows:
\begin{eqnarray}\label{eq:34}
A=\displaystyle\int_{p_1}^{p_2} \mathrm{d}p \int_{x_1}^{x_2} \mathrm{d}x ~ x^2 \left| \psi \right|^2=\left<x^2\right>,
\end{eqnarray}
\begin{align}\label{eq:35}
B&=\int_{p_1}^{p_2} \mathrm{d}p \int_{x_1}^{x_2} \mathrm{d}x \left(\psi x \frac{\partial}{\partial x} \psi^\ast 
+\psi^\ast x \frac{\partial}{\partial x} \psi \right)\nonumber\\
&=\int_{p_1}^{p_2} \mathrm{d}p \int_{x_1}^{x_2} \mathrm{d}x \psi^\ast \left( x \frac{\partial}{\partial x}-\frac{\partial}{\partial x} x \right) \psi\nonumber\\
&=\frac{-1}{\mathrm{i} \hbar}\int_{p_1}^{p_2} \mathrm{d}p \int_{x_1}^{x_2}\mathrm{d}x \psi^\ast \left(x \hat{p}-\hat{p}x \right) \psi=-1,
\end{align}
\begin{align}\label{eq:36}
C&=\int_{p_1}^{p_2} \mathrm{d}p \int_{x_1}^{x_2} \mathrm{d}x \frac{\partial \psi}{\partial x} \frac{\partial \psi^\ast}{\partial x}\nonumber\\
&=-\int_{p_1}^{p_2} \mathrm{d}p \int_{x_1}^{x_2} \mathrm{d}x \psi^\ast \frac{\partial^2 \psi}{\partial x^2}=\frac{\left<p^2\right>}{\hbar^2}.
\end{align}
Here, we took into consideration that the wave functions at the boundary points vanish and the wave function satisfies the normalization condition in phase space,
\begin{equation}\label{eq:37}
\int_{p_1}^{p_2} \mathrm{d}p \int_{x_1}^{x_2} \mathrm{d} x \left| \psi \right|^2=1.
\end{equation}
Hence, it follows that relation
\begin{equation}\label{eq:38}
\frac{\left< x^2 \right> \left< p^2 \right>}{\hbar^2} \geq \frac{1}{4} \Rightarrow \sqrt{\left< x^2 \right>} \cdot \sqrt{\left< p^2 \right>} \geq \frac{\hbar}{2}
\end{equation}
should hold. 
This is the uncertainty relation for position and momentum in the sense of the standard deviation.

The action function is defined with respect to generalized coordinates and momenta, so that the proof of the uncertainty relations naturally extends to the case of angle and angular momentum.
Consequently, the uncertainty relation for the angle and angular momentum operator is equally obtained from commutation relation
\begin{equation*}
\left(\varphi \hat{L}_{\varphi}-\hat{L}_{\varphi}\varphi \right) =-\mathrm{i} \hbar \left(\varphi \frac{\partial }{\partial \varphi}- \frac{\partial}{\partial \varphi} \varphi \right) =\mathrm{i} \hbar.
\end{equation*}
Thus, the uncertainty relation has been generally proved by means of the commutation relation for position and momentum operator in the present formalism.
It is obvious that similar calculation for the position operator $-\mathrm{i}\hbar \displaystyle\frac{\partial }{\partial p}$ and momentum $p$ gives the identical uncertainty relation. 
This is because these operator and observable satisfy commutation relation
\begin{equation*}
\left(p \hat{x}-\hat{x}p \right) =-\mathrm{i} \hbar \left(p \frac{\partial }{\partial p}- \frac{\partial}{\partial p} p \right) =\mathrm{i} \hbar,
\end{equation*}
Evidently, the proof shows the uncertainty relation in the sense of statistical ensemble, which does not reflect the disturbance due to measurement.
Therefore, our description vindicates the statistical formulation of the uncertainty principle.

\subsection{Proof of uncertainty relation for time and energy in terms of commutation relation}
\
\indent It is interesting to prove the uncertainty relation for energy and time by using the commutation relations between them,
\begin{equation}\label{eq:39}
(t\hat{E}-\hat{E}t)=\mathrm{i}\hbar\left(t\frac{\partial}{\partial t}-\frac{\partial}{\partial t}t\right)=-\mathrm{i}\hbar,
\end{equation}
\begin{equation*}
\left(E \hat{t}-\hat{t}E \right) =\mathrm{i} \hbar \left(E \frac{\partial }{\partial E}- \frac{\partial}{\partial E} E \right) =-\mathrm{i} \hbar,
\end{equation*}
where the energy operator and the time operator are represented respectively as
\begin{equation}\label{eq:40}
\hat{E}=\mathrm{i}\hbar \frac{\partial}{\partial t},~~~~~~~~~~~ \hat{t}=\mathrm{i}\hbar \frac{\partial}{\partial E}.
\end{equation}

The probability density in phase space can be considered to be the same as that in energy-time space, assuming that the equivalence relation $\triangle \mathbf{r}\cdot \triangle \mathbf{p} \Leftrightarrow \triangle E\cdot \triangle t$ holds.
For this reason, the integral of the probability density in energy-time space is identical to that in phase space.
Thus, we have good reason for thinking the integral of the probability density in energy-time space as the meaningful one.

To obtain the uncertainty relation for energy and time similar to the case of the momentum operator and position, we perform the following integral in energy-time space.

\begin{equation}\label{eq:41}
\int_{E_1}^{E_2}dE\int_{t_1}^{t_2}dt 
\left(\alpha t \psi^\ast+\frac{\partial \psi^\ast}{\partial t}\right)\left(\alpha t \psi+\frac{\partial \psi}{\partial t}\right)=A\alpha^2+B\alpha+C\geq 0.
\end{equation}
Here,  $E \in[E_1, E_2]$ and $t \in [t_1, t_2]$ represent the domain of variability of energy and time. 
As usual, we may take the variability as $E_1=0, E_2=\infty, t_1=0, t_2=\infty$. At the boundaries the values of the wave function vanish.

From Eq. (\ref{eq:41}), we get
\begin{equation}\label{eq:42}
B^2-4AC\leq 0 \rightarrow B^2/4\leq AC.
\end{equation}
In the next place, we determine $A, B, C$ as follows.
\begin{equation}\label{eq:43}
A=\int_{E_1}^{E_2}dE\int_{t_1}^{t_2}t^2\left|\psi\right|^2dt =\left<t^2\right>,
\end{equation}

\begin{align}\label{eq:44}
B&=\int_{E_1}^{E_2}dE\int_{t_1}^{t_2}dt~ \left(\psi t \frac{\partial \psi^\ast}{\partial t}+\psi^\ast t \frac{\partial \psi}{\partial t}\right)\nonumber\\
&=\int_{E_1}^{E_2}dE\int_{t_1}^{t_2}dt~ \psi^\ast\left(t \frac{\partial }{\partial t}-\frac{\partial }{\partial t}t\right)\psi\nonumber\\
&=\frac{1}{\mathrm{i}\hbar}\int_{E_1}^{E_2}dE\int_{t_1}^{t_2}dt \psi^\ast\left(t \hat{E}-\hat{E}t\right)\psi=-1, 
\end{align}

\begin{align}\label{eq:45}
C&=\int_{E_1}^{E_2}dE\int_{t_1}^{t_2}dt~\frac{\partial \psi}{\partial t}\frac{\partial \psi^\ast}{\partial t}=-\int_{E_1}^{E_2}dE\int_{t_1}^{t_2}dt~\psi^\ast \frac{\partial^2 \psi}{\partial t^2}= \frac{\left<E^2\right>}{\hbar^2},
\end{align}
where we took into account that the wave functions at the boundary points vanish and the wave function satisfies the normalization condition in energy-time space,
\begin{equation}\label{eq:46}
\int_{t_1}^{t_2} \mathrm{d}t \int_{E_1}^{E_2} \mathrm{d} E \left| \psi \right|^2=1.
\end{equation}
This stands for the fact that any particular state $(q, p)$ in phase space is necessarily found over the whole process of time and energy.   
Hence, we obtain
\begin{eqnarray}\label{eq:47}
\frac{\left<E^2\right>\cdot \left< t^2\right>}{\hbar^2}\geq\frac{1}{4},
\end{eqnarray}
namely,
\begin{eqnarray}\label{eq:48}
\sqrt{\left<E^2\right>}\cdot \sqrt{\left< t^2\right>}\geq \frac{\hbar}{2}.
\end{eqnarray}

In doing so, we reach the uncertainty relation for energy and time based on the statistical formulation.

Likewise, we can obtain the uncertainty relation for energy and time, performing the calculation in energy-time space in terms of commutation relation between the energy and time operator,
\begin{equation}\label{eq:49}
(E\hat{t}-\hat{t}E)=\mathrm{i}\hbar\left(E\frac{\partial}{\partial E}-\frac{\partial}{\partial E}E\right)=-\mathrm{i}\hbar.
\end{equation}
From this, it is evident that the same uncertainty relation as Eq. (\ref{eq:48}) is obtained.

Thus, the uncertainty relation not only for position and momentum but also for energy and time has been proved in a general way in terms of commutation relation of operators.
This fact demonstrates that quantum mechanics in phase space provides more general understanding of the uncertainty principle than the configuration-space formulation.

\section{Interpretation of uncertainty relation in terms of action function}\label{AF}
\
\indent Let us examine what the uncertainty relations, (\ref{eq:38}) and  (\ref{eq:48}) mean.
To tell the truth, it shows that the minimum of product of distributions of conjugate coordinate and momentum cannot get smaller than the Planck constant and that the distributions of two canonically conjugate observables are inversely proportional.

However, the information is not satisfactory as such.
If the product of distributions of conjugate coordinate and momentum is larger than the Planck constant, how can we explain the uncertainty relation in a quantitative way?
In fact, relation (\ref{eq:38}) cannot give a satisfactory answer to this question because it is represented as an inequality.
As a question, is it possible that the product of two uncertainties becomes $2.1h$ or $3.4h$ or 20h?
If we understand the uncertainty relation as it is, these three values are all possible, since they are greater than the Planck constant.
In fact, it is a task to clarify which case is true.

The solution to this problem can be given by considering on the basis of the de Broglie relation.
It is significant to show the fact that the contents of the uncertainty relation are sufficiently explained on the basis of the de Broglie relation without using the commutation relation between operators and wave function.
If it is possible, it would be the confirmation that the de Broglie relation is the foundation of the uncertainty relations  and there is another way to explain the relations more reasonably than the previous ones.

The analysis of the de Broglie relation has shown that the action gives the phase of material wave.
Let us consider that starting from relations (\ref{eq:9}) and (\ref{eq:10}), the uncertainty relation obtained in the sense of the standard deviation can be derived directly.

We believe that the nature of quantum lies in the quantization of the action.
In other words, it is in every quantum-physical process that the action should be quantized.
As an example, in three-dimensional case, the action function as an integral along a path is a function in phase space to be written as 
 \begin{equation*}\label{eq:46}
S=\int_{\mathbf{r}_0}^{\mathbf{r}}\mathbf{p}(\mathbf{r}')d\mathbf{r}'=\int_{x_0}^xp(x')dx'+\int_{y_0}^yp(y')dy'+\int_{z_0}^zp(z')dz'=kh,
\end{equation*}
where $k$ is an integer.
For convenience, we shall consider one-dimensional case.
Then the action function is represented as
 \begin{equation}\label{eq:50}
S=\int_{x_0}^xp(x')dx'=nh.
\end{equation}
By the mean value theorem of the integral calculus, we have
\begin{equation}\label{eq:51}
S=\int_{x_0}^xp(x')dx'=(x-x_0)p(\xi)=\triangle x \bar{p},
\end{equation}
where $\xi$ is between $x_0$ and $x$, and $\bar{p}_n$ can be  interpreted as the mean magnitude of momentum. 
Hence, we can write the quantization condition for momentum as
\begin{equation}\label{eq:52}
\triangle x\bar{p}_n=nh.
\end{equation}
Therefore, a quantized momentum in a given interval $\triangle x$ is written as
\begin{equation*}
\bar{p}_n=\frac{nh}{\triangle x}.
\end{equation*}
It is obvious that for $n=1$, $\triangle x$ is the wavelength of material wave.
It means that for a quantum state to be realized, for a given momentum the magnitude of space necessary for a quantum  state at least should be equal to the wavelength of material wave.
If the magnitude of space is less than the wave length, a quantum state is impossible to be realized.

For a given $n$, magnitude of momentum and distribution of coordinate are in the inversely proportional relation.
This shows that the greater momentum is, as a result of wavelength of material wave to get smaller, the weaker the quantization of space is and thus the character of the given system approaches that of classical system characterized by continuum.
For a given $\triangle x$, the smallest variation in momentum due to quantum fluctuation is
\begin{equation}\label{eq:53}
\triangle p=\frac{h}{\triangle x}.
\end{equation}
Therefore, if $\bar{p}_n\gg \triangle p$, then the quantum fluctuation in state of a system is negligibly small and thus the system becomes classical.

On the other hand, if $\bar{p}_n\approx \triangle p$, the quantum fluctuation proves to cause remarkable change in state of a particle and therefore its state becomes quantum.
The influence of $\triangle p$ due to quantum fluctuation on the velocity of a particle  is the criterion for distinguishing between microscopic particle and macroscopic one.
For a quantum fluctuation in momentum, $\triangle p$, on account of relative difference between masses of macroscopic particle and microscopic one the change in velocity of macroscopic particle is considerably small compared to that of microscopic particle and as a result the macroscopic particle is insensitive to quantum fluctuation.

On the other hand, we can consider the uncertainty relation in the sense of the standard deviations of canonically conjugate quantities.

Our view is that quantum fluctuation puts in action a family of virtual trajectories, i.e., an ensemble of trajectories around a classical trajectory and thus it is possible to imagine the deviation of actions of quantum trajectories from the action of the classical trajectory.

The action integral for the classical path usually has a minimum according to the principle of least action.
For the classical path we have as a minimum
\begin{equation}\label{eq:54}
S_0=\int_{x_0}^xp_0(x')dx'=\triangle x \bar{p}_0=n_0h,
\end{equation}
where $n_0$ is an integer.
For quantum processes taking place in the equal interval, actions are represented as
\begin{equation}\label{eq:55}
S_n=\int_{x_0}^xp_n(x')dx'=\triangle x \bar{p}_n=nh,
\end{equation}
where $n$ is an integer.

The subtraction of Eq. (\ref{eq:54}) from Eq. (\ref{eq:55}) determines the quantum fluctuation as
\begin{equation}\label{eq:56}
|S_n-S_0|=\triangle S_n=\triangle x |\bar{p}_n-\bar{p}_0|=\triangle x \triangle \bar{p}_n=|n-n_0|h.
\end{equation}

From Eq. (\ref{eq:56}), we have as the standard deviation of action
\begin{equation*}
\triangle S=\sum_nW_n|S_n-S_0|=\triangle x\sum_nW_n\triangle\bar{p}_n,
\end{equation*}
where $W_n$ refers to the probability of a fluctuating quantum path.

Equating $\sum_nW_n\triangle\bar{p}_n$ with the mean value of momentum deviation, $\triangle p$, we have
\begin{equation}\label{eq:57}
\triangle x\sum_nW_n\triangle\bar{p}_n=\triangle x\cdot \triangle p=kh,
\end{equation}
where $k$ is an integer.

From Eq. (\ref{eq:57}), it follows that the minimum condition of quantum fluctuation is
\begin{equation}\label{eq:58}
\triangle x\cdot \triangle p=h.
\end{equation}
Since there is no quantum fluctuation in momentum in the case of free particle of a definite momentum, Eq. (\ref{eq:57}) reads 
\begin{equation}\label{eq:59}
\triangle x\cdot \triangle p=0,
\end{equation}
which should be considered to represent a classical state free of quantum fluctuation.
Eqs. (\ref{eq:57})--(\ref{eq:59}) imply that the uncertainty can change exactly by an integral multiple of the definite value, $h$, which integral multiple begins from zero.

Thus, as the essential contents of the uncertainty relation, the facts are explained satisfactorily in terms of the action function in phase space that  there exists the quantum of phase space, $h$ and the uncertainty relation is determined by an integral multiple of $h$.
As an important understanding of classical limit of this principle, it is satisfactorily explained also the fact that for a particle of great momentum the uncertainty of position due to quantum fluctuation is small.
This method explains more than that proving the uncertainty relation in terms of the commutation relation of operators in the sense of standard deviation.
In fact, the relation of Eq. (\ref{eq:52}) is not possible to be expounded correctly in the sense of the standard deviation. 

Next, let us consider the uncertainty relation for time and energy.
Quantum states in energy-time space are represented  by action as
\begin{equation}\label{eq:60}
S=\int_{t_0}^tE(t')dt'=nh.
\end{equation}

According to the mean value theorem of the integral calculus,  the above expression can be recast as
\begin{equation}\label{eq:61}
S=\int_{t_0}^tE(t')dt'=(t-t_0)E(\tau)=\triangle t \bar{E}=nh,
\end{equation}
where $\tau$ is between $t_0$ and $t$ and $\triangle t$ is a duration of quantum process.
Therefore, the relation between the magnitude of energy and the duration of time is obtained as
\begin{equation}\label{eq:62}
\bar{E}=\frac{nh}{\triangle t},
\end{equation}
\begin{equation}\label{eq:63}
\triangle t=\frac{nh}{\bar{E}}.
\end{equation}
If $n$ is given, the magnitude of energy and the duration of time are inversely proportional to each other.
This indicates that the greater the energy is, the weaker the quantization of time, and as a result, the character of a system approaches that of classical system assuming continuous time. 

As in the case of dealing with the uncertainty relation for position and momentum, for classical and quantum processes we determine the deviation of action as
\begin{equation}\label{eq:64}
|S_n-S_0|=\triangle S_n=\triangle t \left|\bar{E}_n-\bar{E}_0\right|=\triangle t \cdot \triangle \bar{E}_n=|n-n_0|h.
\end{equation}
With the help of Eq. (\ref{eq:64}), the standard deviation of action is written as
\begin{equation*}\label{eq:61}
\triangle S=\sum_nW_n|S_n-S_0|=\triangle t\sum_nW_n\triangle\bar{E}_n.
\end{equation*}
Since $\sum_nW_n\triangle\bar{E}_n$ is the mean value of energy deviation, $\triangle E$, it follows that
\begin{equation}\label{eq:65}
\triangle S=\triangle t\cdot \triangle E=kh,
\end{equation}
where $k$ is an integer.

Naturally, from Eq. (\ref{eq:65}), as the minimum condition of quantum fluctuation, we get
\begin{equation}\label{eq:66}
\triangle t \cdot \triangle E=h.
\end{equation}
For a definite energy, since quantum fluctuation vanishes, Eq. (\ref{eq:65}) is written as
\begin{equation}\label{eq:67}
\triangle t\cdot \triangle E=0.
\end{equation}

It is essential that from Eqs. (\ref{eq:9}) and (\ref{eq:10}) issue Eqs. (\ref{eq:38}) and (\ref{eq:48}).
Therefore, it would be correct to start with Eqs. (\ref{eq:9}) and (\ref{eq:10}) in  order to establish the foundation of the uncertainty relation.
Obviously, this interpretation is different from  Heisenberg's interpretation of the uncertainty relation, since it is not related to measurement.  

It is important to argue the lowest limit of uncertainty.
The matter is whether it is $h$ or $ \hbar $ or $\hbar /2$.
It is reasonable to take the limit as $h$ in that the foundation of the uncertainty relation is the de Broglie relation and the direct explanation of the uncertainty relation in terms of the action function is more straightforward and general than that in terms of standard deviation.
In fact, we take the cell in phase space as $h$ in statistical mechanics.
Ultimately, we can conclude that the explanation of the uncertainty relation in terms of the action function is more correct and general than the previous, and contains the complete contents of the uncertainty principle.

\section{Results and discussion}\label{sec:Discussion}
\
\indent The perspective of this formalism leads to an alternative understanding of why the uncertainty relations exist and how to prove them in a general way.

Obviously, once we argue the uncertainty principle in the sense of the standard deviation, the foundation of the uncertainty principle is relations (\ref{eq:9}), (\ref{eq:10}) given by the de Broglie relation.
These relations are another representation of the de Broglie relation as well.
Without the de Broglie relation, it is impossible to imagine neither the form of the wave function and quantum operators  nor the uncertainty relations.
As a matter of fact, we started from the de Broglie relation to arrive at the uncertainty relations.
Therefore, the de Broglie relation evidently is the foundation of the uncertainty principle and the latter is nothing but a corollary of the former.
In this sense, it is logic to consider the uncertainty relation not to have the status of principle.
Relations (\ref{eq:9} and  (\ref{eq:10} become the foundation for expounding the contents of the uncertainty relation perfectly.
The lager the momentum of a particle is, the smaller the size of the region in position space required for a quantum state of the particle. 
This is because in this case the wavelength of material wave gets shorter.
This relation seems to be different from the Heisenberg relation $\triangle x \cdot \triangle p \geq \hbar$, but actually represents the essence of the indeterminacy of  observables in quantum processes in a more intuitive way  compared to the latter.

It is necessary to review the uncertainty principle in more details.
The argument below is the summarized citation of Home and Whitaker's description and our additions\cite{Home}.
It is important to note that the derivation of the uncertainty principle uses no input from quantum dynamics. 
Even if one uses a wave function having the wrong symmetry and violating the Schr\"{o}dinger equation, the uncertainty relation will be necessarily established. 
The uncertainty principle is thus insensitive to any modification of the Schr\"{o}dinger equation.
However, the uncertainty principle definitely depends on the commutation relation between a canonical variable and the operator for the corresponding conjugate variable.
As mentioned above, the identical commutation relation between canonically conjugate variable and operator yields one and the same uncertainty relation.
It should be emphasized that our generalized proof of the uncertainty relations demonstrates how to obtain the law of distributions of canonically conjugate observables in terms of the very commutation relation.
To demonstrate the uncertainty relations, the joint measurement of canonically conjugate observables is a necessity rather than a prohibition.
In fact, this principle reflects only the reciprocal relation of distributions between two canonically conjugate observables in case they are jointly measured. 
If the simultaneous measurement of canonically conjugate observables were impossible, there could not be the proof of the uncertainty relations by means of the method of determining the statistical distribution of them as well.   

The interpretational significance of the uncertainty principle may be stated in one of the two following ways. 

The first approach is based on Heisenberg's thought experiments.
Its tenor is that there is the reciprocal relationship between  distributions of errors obtained in measurement of canonically conjugate variables due to intervention of measurement.
In  the end, this interpretation places the center of the  principle in the disturbance due to measurement rather than in objective property of quantum state.

Heisenberg's perspective on the  uncertainty principle has been quite widely held right up to the present day. 
Due to this interpretation of the uncertainty principle, quantum mechanics was doomed to choose either configuration space or momentum space, thus being relegated to a narrow space imperfect for consistent quantum theory.
As shown in Sec. \ref{Possibility}, the simultaneous determination of position and momentum is necessary and possible also for quantum theory.
In fact, it is untenable to connect the uncertainty relations with measurement, since there are the interaction-free measurements as well \cite{Renninger}. 
It is important to pay attention to the fact that the wave function and operators are not implicated in human's behavior for measurement and have nothing to do with measurement itself, while the proofs of the uncertainty relations are offered by nothing but the wave function and operators.
This shows that the intervention of measurement responsible for the uncertainty principle has no scientific justification.

The second approach to interpreting the uncertainty principle  states that uncertainty in the value of a dynamical variable is referred to the statistical spread over the measured values for various identical members of the ensemble of systems. 

The operational significances of the two approaches are fundamentally different from each other. 
In the second, there is no question in making simultaneous measurement of the dynamical variables related to a single particle.
As it is, this concept embraces the essence of the first approach too, where the uncertainty in a single measurement is interpreted as the imprecision in the measured value of a dynamical variable for a single particle rather than the stochastic character inherent to quantum states. 
This is because if there were not the possibility of joint measurement,  there would not be the calculation of the standard deviation in the sense of joint distribution, or the variance reflecting an objective quantum state of a microscopic particle.  
For this reason, it is reasonable to refer to the uncertainty as statistical distribution. 
According to the statistical interpretation of the uncertainty principle, we come to know that in principle, it is consistent to use phase space for the studies of quantum mechanics. 

It is obvious that Heisenberg's idea in the original form should certainly not be regarded as providing a proof of the uncertainty principle and being a foundation of quantum mechanics.
As already shown, the uncertainty principle, in essence, is not a principle but a provable corollary which is derived from a deeper principle, specifically the de Broglie relation. 

Our approach can adduce good reason to support the second approach.
The fundamental equation of our formalism is the quantum version of Liouville's equation, and therefore the uncertainty principle proves to be a logical corollary of the present formalism \cite{Jongchol}.
According to Louville's theorem, the consideration of the density of phase points forming a statistical ensemble along a phase trajectory indicates that phase volume occupied by it remains constant. 
It turns out that the volume in configuration space occupied by a statistical ensemble of microscopic particles and that in momentum space are inversely proportional to each other. 
Hence, we may state in the language of the uncertainty relation that the higher the accuracy of position measurement is, the lower that of momentum measurement. 
Exactly representing, the wider the distribution of position is, the more narrow that of momentum. 
This argument can be regarded as an account of the uncertainty relation which issues from the  point of view of the statistical formulation. 

In this connection, what is most important for quantization is the existence of the quantum of both phase space and energy-time space, i.e., $h$ rather  than the uncertainty relation which is proved by the wave function and operator given by the procedure: Action $\rightarrow$ Form of wave function $\rightarrow$ Operator $\rightarrow$ Commutation relation $\rightarrow$ Uncertainty relation.
In fact, the Schr\"{o}dinger equation is not involved in the first approach to interpreting the uncertainty principle.
It should be noted that the Schr\"{o}dinger equation had been discovered before the proposition of the uncertainty principle and is not underpinned by the principle.
Should we not introduce the one-to-one correspondence between position and momentum according to scheme, (\ref{eq:1}), we could not obtain the Schr\"{o}dinger equation.   
Our discussion gives answer to the complementarity principle as well.
In fact, it is impossible to interpret in a quantitative way the uncertainty principle on the basis of the complementarity principle which is characterized by philosophical issues.
Einstein rejected the quantum philosophy of Copenhagen to make this relation a superstructure of quantum mechanics without proof.
The history of the development of quantum mechanics has witnessed a lot of attempts to prove this relation in a general way.
It is Heisenberg itself as the initiator of this principle who endeavored to prove it in a general way.
As it is, we are bound to note the logic consequence that if it were proved, then the principle would be no longer a principle.

All these arguments lead to the conclusion that it is possible to prove the uncertainty principle in a deductive way based on the de Broglie relation.
Also, it should be emphasized that the interpretation of the uncertainty relation in terms of the action function is more essential and intuitive than that in terms of the product of standard deviations of canonically conjugate variables.
In fact, the quantization condition of action yields all the ingredients of the uncertainty principle.
Our investigation shows that the action is the foundation of the uncertainty principle and the uncertainty relations issue naturally from the quantization of the action.
Eventually, we can confirm that the de Broglie relation performs a pivotal role in formulating the quantum theory.

\section{Conclusion}\label{concl}
\
\indent We have investigated an alternative formulation of the uncertainty principle based on the statistical formalism of  quantum mechanics built by our previous paper \cite{Jongchol}. 

Our work has shown that it is possible to establish the uncertainty principle as a provable quantum rule on the basis of a more profound principle. 
As shown above, the de Broglie relation enables us to assume the form of the wave function which is defined necessarily in phase space.
Furthermore, the dynamical structure of the wave function in terms of the action function leads to finding quantum operators to apply to the wave function in order to determine dynamical quantities,  thus establishing the complete system of quantum operators.
The system of quantum operators enables us to prove the uncertain relations in a general way in terms of commutation relation.
Actually, the commutation relations between quantum operators naturally lead to the generalized proof of the uncertainty relations.

Such a methodology of inferring the uncertainty principle is given by the statistical formalism of quantum mechanics.  
This formalism produces within its framework the fundamental equation without recourse to the other formulations of quantum mechanics. 
On the other hand, it offers the explanation of why to introduce quantum operators and how to find out quantum operators corresponding to dynamical quantities.
Obviously, this formalism demonstrates that the action function is the foundation providing the complete system of quantum operators. 

Meanwhile, our work has shown that it is possible to expound in a direct way the uncertainty relation in terms of the action function which the de Broglie relation suggests, without using the wave function and commutation relation of operators. 
All the contents of the uncertainty principle actually can be derived from the quantized action.

In the end, it is a matter of course that we reach the conclusion that the uncertainty principle as the dictum of the impossibility of simultaneous measurement of canonically conjugate observables should be reassessed from the point of view of the foundation and interpretation of quantum mechanics. 

In conclusion, our work confirms that the formalism of quantization in terms of statistical ensemble in phase space is consistent with the fundamentals of quantum mechanics, and offers a possibility of resolving some open questions including the uncertainty principle that still remain unsolved.
  
\section*{Acknowledgments}
We thank Profs. Ji-Min Pak, Won-Chol Ri and Il-Yong Kang from Kim Chaek University of Technology for discussion and help. 

\end{document}